# On the first-order structural transition in the multiferroic hybrid organic-inorganic perovskite-like formate $[(CH_3)_2NH_2][Mn(HCOO)_3]$

M. Sánchez-Andújar[a], L. C. Gómez-Aguirre[a], B. Pato Doldán[a], S. Yáñez-Vilar[a], R. Artiaga[b], A. L. Llamas-Saiz[c], R. S. Manna[d], F. Schnelle[d], M. Lang[d], F. Ritter[d], A. A. Haghighirad[d,e], M. A. Señarís-Rodríguez[a*]

[a] Department of Fundamental Chemistry, University of A Coruña, 15071 A Coruña, Spain.
[b] Department of Industrial Engineering II, University of A Coruña, Campus de Esteiro, Ferrol 15403, Spain.
[c] RIAIDT X-Ray Unit, University of Santiago de Compostela, 15782 Santiago de Compostela, Spain
[d] Institute of Physics, Goethe University, 60438 Frankfurt am Main, Germany
[e] Clarendon Laboratory, University of Oxford, Parks Road, Oxford OX1 3PU, U.K.

**Abstract**

In this work we explore the overall structural behaviour of the $[(CH_3)_2NH_2][Mn(HCOO)_3]$ multiferroic compound across the temperature range where its ferroelectric transition takes place by means of calorimetry, thermal expansion measurements and variable temperature powder and single crystal X-ray diffraction. The results clearly proof the presence of structural phase transition at $T_t$ ~187 K (temperature at which the dielectric transition occurs) that involves a symmetry change from R-3c to Cc, twinning of the crystals, a discontinuous variation of the unit cell parameters and unit cell volume, and a sharp first-order-like anomaly in the thermal expansion. In addition, the calorimetric results show a 3-fold order-disorder transition. The calculated pressure dependence of the transition temperature is rather large ($dT_t/dP$ = 4.6 ± 0.1 K/kbar), so that it should be feasible to shift it to room temperature using adequate thermodynamic conditions, for instance by application of external pressure.

**Introduction**

Multiferroic materials, in which magnetic and electrical ordering coexist,[1] have received much attention in the last few years in view of their potential applications to magnetic storage, novel circuits, sensors, microwave and high-power applications.[2,3]

However, up to date single phase multiferroics are rare as the conventional mechanism for ferroelectric and cooperative magnetism are mutually exclusive.[4] Beyond transition-metal oxides, that have been extensively investigated,[5] the family of hybrid inorganic-organic materials are attracting increased attention.[6-9]

Among them, the dense metal–organic frameworks (MOF) of formula $[(CH_3)_2NH_2][M(HCOO)_3]$ (with $M^{2+}$ = $Mn^{2+}$, $Co^{2+}$, $Ni^{2+}$ and $Fe^{2+}$) [7,10] with perovskite $ABX_3$ architecture (where A = $[(CH_3)_2NH_2]^+$, B = $M^{2+}$, X = $HCOO^-$) are specially interesting as type I-multiferroics: they combine structural flexibility, relatively good stability, low density and cooperative magnetic and electrical properties.

These latter arise from different active "subsystems" (Figure 1):

i) A rather robust framework based on $[BO_6]$ octahedra, which are connected through the short organic formate ligands, and that displays cooperative magnetism at low temperatures ($T_C$: 8-36 K). [7,11,12]

ii) Dipolar dimethylammonium cations (DMA), $[(CH_3)_2NH_2]^+$, that occupy the A-sites of the perovskite architecture, are linked to the framework via hydrogen bonds, and experience an order/disorder process.[10] This latter seems to be the underlying cause for the dielectric transition, which appears at $T_t$ in the range 165-190 K, depending on the transition metal (TM) ion $M^{2+}$ ($M^{2+}$ = $Mn^{2+}$, $Co^{2+}$, $Ni^{2+}$, $Fe^{2+}$ and $Zn^{2+}$).[7] Such dielectric transition, that was initially considered of paraelectric–antiferroelectric type [7,13] has more recently been associated to improper ferroelectricity.[14] Furthermore, at the same temperature $T_t$ a phase transition occurs, that was considered to be of second order.[7]

In view of the different origins of the magnetic and ferroelectric order, weak direct magnetoelectric (ME) coupling should be expected in these systems, even if very recent publications are highlighting the potential of alternative indirect coupling mechanisms for obtaining much larger ME effects in these and related MOFs.[15-17]

Out of all [(CH$_3$)$_2$NH$_2$][M(HCOO)$_3$] compounds the Mn-formate -which shows a magnetic transition at $T_N \approx 8.5$ K,[7,11,12] a dielectric transition at $T_t \approx 190$ K [7,10] and according to a very recent report a new magnetoelectric coupling in the paramagnetic state near the ferroelectric transition [17]- is so far the best characterized one,[11,12] even if a number of questions remain open, for example, if the structural transition is a first or second order transition.

According to the literature, at room temperature this compound shows trigonal symmetry (space group: *R*-3*c*, Z = 6), Fig. 1a. In this phase, the [(CH$_3$)$_2$NH$_2$]$^+$ (DMA) cations are disordered with the nitrogen atom equally distributed in three different positions, and the Mn$^{2+}$ ions residing in a regular [MnO$_6$] octahedral environment. Meanwhile at 100 K (the only temperature below the dielectric transition for which crystal data are available [10]) this compound shows monoclinic symmetry (space group: *Cc*, Z = 4), where the DMA cations are cooperatively ordered and the [Mn(HCOO)$_3$]$^-$ framework is strongly distorted,[10] see Fig. 1b.

Based on the existing data, it could be proposed that the driving force for the dielectric transition is either the ordering of the hydrogen atoms of the DMA cations, [7,13] similarly to what occurs in KDP-compounds [18] and/or the cooperative freezing of the molecular rotation of the DMA cations, [10] even if a direct and unequivocally proof at the transition still remains elusive.

These mechanisms of order-disorder phase transition have subsequently been invoked and extended to explain the dielectric response observed in other hybrid inorganic-organic materials, such as for example $[(CH_3)_2NH_2]_2[KCo(CN)_6]$, [19] $[(CH_3)_2NH_2]_n[Fe^{III}Fe^{II}(HCOO)_6]_n$, [20] $(NH_4)M(HCOO)_3$ with ($M^{2+}$ = $Mn^{2+}$, $Co^{2+}$, $Ni^{2+}$, $Fe^{2+}$ and $Zn^{2+}$) [21] and $(HIm)_2[KFe(CN)_6]$ (HIm = $C_3N_2H_5^+$ protonated imidazole) [22] with the focus placed on the role of the alkyl/aryl-ammonium cations.

The goal of this paper is to shed more light on the phase transition experienced by the Mn-formate $[(CH_3)_2NH_2][Mn(HCOO)_3]$ that, as indicated above, is associated with the dielectric transition and the onset of magnetoelectric coupling in the paramagnetic state.[17]

For this purpose, and at difference with the previous works, we explore the overall structural behaviour of the $[(CH_3)_2NH_2][Mn(HCOO)_3]$ compound across the temperature range where the dielectric transition takes place by means of high-resolution thermal expansion measurements on high-quality single crystals, calorimetric experiments on crushed single crystals and temperature dependent X-ray diffraction studies on both powders and single crystals.

**Results and Discussion**

**Differential scanning calorimetry**

The obtained curves, that are fully reproducible and very clean, reveal an endothermic transition that is observed upon heating around 190 K, temperature at which the dielectric transition occurs. [10] On cooling the corresponding exothermic transition is seen at around 180 K (see Fig. 2).

The observed large thermal hysteresis, which increases with the faster heating and cooling rates (see Fig. 2), implies that the observed transition is of first-order character, as thermal hysteresis is thermodynamically forbidden in a second-order transition. [23]

The changes in the enthalpy $\Delta H$ (J/mol) and the entropy $\Delta S$ (J/mol K) were determined from the area under the heat flow/time curves. The results, which depend on the measuring conditions as shown in Table S1 of Electronic Supplementary Information (ESI), yield an average $\Delta H_{av} \sim$ 1673 (endo) - 1708 (exo) J/mol and an average $\Delta S_{av} \sim$ 8.8 (endo) - 9.4 (exo) J/mol K for this phase transition. Taking into account that for an order-disorder transition $\Delta S = R \ln(N)$, where R is the gas constant and N is the ratio of the number of configurations in the disordered and ordered system, a value of N 3 is calculated, thus pointing out towards a 3-fold order-disorder model for the DMA cation.

These values are higher, even though of the same order, than those recently reported for the related deuterated Co-compound [(CD$_3$)$_2$ND$_2$][Co(DCOO)$_3$] , with N ~ 1.5, where the transition leads to a partially-ordered LT phase, as confirmed by single crystal X-ray diffraction studies; [24] or for the Mg-formate [(CH$_3$)$_2$NH$_2$][Mg(HCOO)$_3$], N ~ 1.7, where the transition takes place near room temperature and is much broader. [25]

On the other hand, they are almost one order of magnitude higher than those initially measured by Jain et al. on single crystals of [(CH$_3$)$_2$NH$_2$]M(HCOO)$_3$ (M$^{2+}$ = Mn$^{2+}$, Co$^{2+}$, Ni$^{2+}$, Fe$^{2+}$ and Zn$^{2+}$) that were nevertheless prepared using slightly different reagents and different synthetic conditions and whose low temperature crystal structures could not be solved due to twin formation. [7,13]

As for the kinetics of the change of phase, we have estimated the activation energy from the obtained DSC data using the Kissinger equation: [26]

$$E_a \beta (RT_p^2) = A e^{-E_a/RT_p}$$ (Eq.1)

where A is the frequency factor, $T_p$ is the peak temperature, R is the gas constant and $\beta$ is the heating rate, which is expressed as $\beta = dT/dt$.

The activation energies $E_a$ calculated for the endothermic and exothermic processes are 209 and 159 kJ/mol, respectively. The fact that these values are very close to each other indicates a very similar mechanism for the phase change that occurs upon cooling and heating, at variance with the case of the endothermic/exothermic transition experienced by the corresponding Co-formate. [27]

**Powder X-ray diffraction**

Powder X-ray diffraction (PXRD) experiments performed at different temperatures revealed that this compound experiences a rather abrupt structural phase transition on cooling below 185 K.

As shown in Figure 3, for $T \geq 190$ K the only phase present is the one with $R\text{-}3c$ symmetry (the HT structure displayed in Fig. 1a). At 185 K this phase coexists with small amounts of a low-temperature phase that becomes the only one present for $T <$ 185 K. This phase displays $Cc$ symmetry (the LT structure displayed in Fig.1b).

From Le Bail fit to the PXRD data (see Fig. S1 of ESI) obtained at different temperatures the cell parameters for the HT and LT-phases were calculated. The results are presented in Figure S2 of ESI, in terms of the original trigonal (HT) and monoclinic (LT) cells.

In addition, to facilitate their direct comparison it was analyzed how the unit cell of the HT $R$-$3c$ structure transforms into the LT $Cc$ using the single crystal X-ray crystallographic information provided in reference [10], finding the following vector relationships:

$$\vec{a}_m = -\vec{a}_h - 2\vec{b}_h; \quad \vec{b}_m = -\vec{a}_h; \quad \vec{c}_m = \tfrac{1}{3}\vec{a}_h + \tfrac{2}{3}\vec{b}_h - \tfrac{1}{3}\vec{c}_h$$

Figure 4 shows the calculated HT-phase unit cell parameters (expressed in terms of the monoclinic cell) together with those of the LT-phase, as well as their temperature dependence.

A general trend is observed in Fig. 4 regarding the length of the unit cell parameters of both the LT and HT-phases that increase with temperature, as expected. This effect is more pronounced in the thermal expansion data for $a_{LT}$ and $c_{HT}$ (Fig. 5).

Remarkably, a discontinuous variation of the cell parameters occurs at $T_t$, with $c_m$ experiencing the largest expansion ($\Delta c_m/c_m \sim 1.1\% > \Delta a_m/a_m \sim 0.7\% > \Delta b_m/b_m \sim 0.08\%$) and $\beta$ abruptly increasing by 1.5 % upon warming.

As for the cell volume, it also experiences a small but clear jump at $T_t$, $\Delta V/V \sim 0.2\%$ (see Fig. S3 of ESI).

Unfortunately, no further structural details could be obtained from the PXRD data due to marked preferential orientation that prevented from obtaining good enough Rietveld refinements.

**Thermal expansion**

To study the thermal expansion of $[(CH_3)_2NH_2][Mn(HCOO)_3]$ across the transition a single crystal of a suitable size (1.00 x 0.80 x 0.75 mm$^3$), see inset Fig. 5, was mounted on the $(1\text{-}12)_{HT}$ oriented crystal facet.

Fig. 5 shows a very sharp (transition width $\sim$ 0.5 K) and large step-like variation in the relative length change $\Delta l(T)/l = [l(T) - l(T_0)]/l$ with $T_0 = 4.2$ K at 188.5 K, demonstrating the first-order nature of the phase transition. This transition temperature is very close to that of the dielectric transition.

The discontinuous length change at the transition, $\delta l/l \sim -5 \times 10^{-3}$, is about five times bigger than in the case of the H-bonded potassium hydrogen phosphate (KTP) [28] or in BaTiO$_3$, [29] which also experiences a first-order phase transition at $T_C$.

It is worth mentioning that compared to other MOFs that display large thermal expansion coefficients, [30] the thermal expansion anomaly observed here, i.e. a strong contraction/expansion at the phase transition, is not related to the elimination/insertion of new guest species or gases from the material, but originates from the order-disorder process of the DMA cations inside the cavities and the concomitant deformations of the

[Mn(HCOO)$_3$]$^-$ framework (see below), that in turn are directly connected to the dielectric transition.

Another interesting aspect is that despite the sharpness of the abrupt length change at $T_t$, this hybrid inorganic-organic crystal does not crack crossing $T_t$ and the experiment could be repeated several times.

**Single crystal X-ray diffraction**

Further studies around the transition temperature were carried out by means of single crystal X-ray diffraction.

The more relevant results are the following:

(a) In these experiments the structural phase transformation from *R-3c* to *Cc* takes place at ~ 187 K. The exact temperature depends on whether the experiment is performed on cooling or heating as well as on the rate of the experiment. This process is very fast, reversible and can be repeated many times without cracking or alteration of the crystal (see Fig. S4 of ESI).

(b) Concomitant to the transformation from *R-3c* to *Cc* the crystal gets twinned. This process is reversible so that the twinned crystal (*Cc* LT-phase) becomes again a single grain (*R-3c* HT-phase) upon heating. It is worth noting that the presence of twins and multi-domain formation is very common in ferroelectric mixed oxides with perovskite structure and a first-order phase transition (such as PbTiO$_3$, BaTiO$_3$, etc).[31]

In this context, in the case of the single crystal studied here (with size ~ 0.21 x 0.28 x 0.60 mm$^3$) the successful refinement of the crystal data obtained at $T$ =185 K, revealed that the monoclinic LT-phase contains three twin domains with relative proportions: 60.8(1) %, 22.9(1) % and 16.3(1) %.

The relative orientation of these twin domains could be sorted out from indexing of all reflections carefully with I/σ(I) > 8 selected from the 983 collected diffraction images (see experimental section). The presence of the three twin domains can also be observed in the pseudo-precession photographs generated from the collected diffraction images in different selected reciprocal space planes ([0kl], [h0l], [hk0], see Fig. 6 and S5 of ESI), this latter being particularly revealing as it shows several extra spots along the $c^*$ axis direction, while it does not show any extra spots in the [hk0] plane.

This implies that in reciprocal space the three twin domains are related to each other by a rotation of approximately 120 degrees around crystallographic axis $c^*$ (axis that is common to both the trigonal and monoclinic reciprocal cells) that in turn is coincident with the $c$- axis of the direct trigonal cell.

Each of these twin domains corresponds in fact to the ordering of the nitrogen (N) atom of the DMA cation into one of the three possible orientations inside the cage formed by the anion framework in the HT trigonal phase (see Fig. S6 of ESI).

This finding supports the mechanism of cooperative freezing of the molecular rotation of the DMA cations that we have proposed in [10] in order to explain the origin of the associated dielectric transition in this compound. Furthermore, in the trigonal phase (for $T > T_t$) the DMA cations would not be static but rotating inside the $[Mn(HCOO_3)]^-$ framework around an axis that passes through the two carbon (C) atoms of the DMA and is parallel to the $c$-axis. This gives rise to the 3-fold disorder of the N atoms of the DMA cations observed in the X-ray diffraction experiments and to the paraelectric behavior. Once the molecular rotation of the DMA cations freezes around 186 K and the DMA cations get their positions fixed inside the cavities in the monoclinic phase, each of the three minima of the well potential would get favoured giving rise to the formation of the twin domains found here. This cooperative order of the DMA cations and their associate dipoles are the ones that give rise to the electrically ordered low-temperature phase, as also observed very recently in the cases of the $[(CH_3)_2NH_2]_2[KCo(CN)_6]$ [19] and $[(CH_3)_2NH_2]_n[Fe^{III}Fe^{II}(HCOO)_6]_n$ [20] compounds.

(c) From the resolution of the crystal structure at temperatures just above and below the transition temperature, namely 190 K and 185 K, relevant structural information could be obtained across the phase transition which is summarized in Table I and Table S2 of ESI.

As it can be seen there, this structural transition gives rise to important changes not only in the DMA cations but also in the $[Mn(HCOO)_3]^-$ framework.

As for the DMA cations, besides the already mentioned disorder-order process of the N-atoms on cooling through $T_t$, significant variations occur in their C-N bond length (Table I) and C-N-C bond angle (Table S2). In this context, the observed increase in the C-N bond length and the associated closing of the C-N-C bond angle in the LT-monoclinic phase are in fact due to the weakening of the intra-DMA H-N bond as a consequence of the strengthening of the H-bond between the DMA and the framework.

In addition, the $[Mn(HCOO)_3]^-$ framework gets significantly distorted as compared to the situation found in the HT trigonal phase, where the $Mn^{2+}$ cations show a regular octahedral environment and both C-O distances and the O-C-O angles are identical for all the formate ions (see Table I).

On cooling below $T_t$ the environment of the $Mn^{2+}$ gets distorted with six different Mn-O distances (four shorter and two longer compared to the HT-phase) and the $[MnO_6]$ octahedra display three different torsion angles. On the other hand, six slightly different C-O distances and three O-C-O angles are detected for the formate anions (see Table I).

Another interesting result is the variation in the [MnO$_6$] octahedral tilting i.e., from a$^-$ a$^-$ a$^-$ (in Glazer tilt system) [32] in the HT-trigonal phase to a$^-$ b$^-$ c$^-$ in the LT-monoclinic phase (Table I).

These structural data reveal that the framework of the [(CH$_3$)$_2$NH$_2$][Mn(HCOO)$_3$] compound is rather flexible, and experiences several distortions (in terms of changes in bond angles and bond lengths) in response to the order/disorder transition of DMA cations.

This phenomenon of structural "adaptation" is very well known in mixed oxides with perovskite structure ABO$_3$ where the [BO$_6$] octahedra can display different tilts, distortions, etc. in order to alter the volume of the resulting central cavity so as to adapt it to the radius of the A cation. [33]

In the case of the here studied perovskite-like formate [(CH$_3$)$_2$NH$_2$][Mn(HCOO)$_3$] the tilting of the [MnO$_6$] octahedra also seems to play a significant role in tuning the size and shape of the resulting pseudo-cubooctahedral cavities. Here, the tilting of the octahedra in the HT-phase leads to a cavity with higher symmetry that is adequate to host the disordered DMA cation. In turn, the rotation of the [MnO$_6$] octahedra on crossing $T_t$ provokes that the cavity shrinks along two of the three 2-fold axes perpendicular to the trigonal 3-fold axis in $R$-$3c$ while it expands along the third one that remains as the 2-fold axis in the monoclinic lattice of LT phase. The result is a cavity with a size and shape more adequate to host the ordered DMA cation.

We propose that both the cooperative ordering of the DMA-cations together with the presence of a flexible [Mn(HCOO$_3$)]$^-$ framework are basic "ingredients" for the appearance of sharp dielectric transitions in this MOF and related materials based on order/disorder of A-site cations.

Finally, as the interest of this material would greatly increase if the transition could be shifted to room temperature, we have evaluated how sensitive it is to the application of pressure.

Since the observed transition is of first order, its pressure dependence is represented by the Clausius-Clapeyron equation:

$$\frac{\partial T_t}{\partial p} = \frac{T_t (\Delta V / V) V_{mol}}{\Delta H_{molar}} \qquad \text{(Eq.2)}$$

Where $T_t$ is the transition temperature, $\Delta V/V$ is the relative volume change at $T_t$, $V_{molar}$ is the molar unit cell volume and $\Delta H_{molar}$ represents the change in the molar enthalpy.

$\Delta V/V$ at the transition was taken from the single crystal results ($\Delta V/V \approx 0.3 \times 10^{-2}$) and $\Delta H_{molar}$ was taken from the DSC-data (1700 Jmol$^{-1}$). $V_{molar}$ (1387.64 × 10$^{-7}$ m$^3$/mole) was calculated for the HT-phase from the single crystal data at 190 K ($V_{molar} = N_A V_{cell}/Z$, where $N_A$ is the Avogadro number, $V_{cell}$ is the unit cell volume and Z is the number of structural units per unit cell)

From the obtained data the following pressure dependence of $T_c$ was calculated:

$dT_c/dP = (4.6 \pm 0.1)$ K/kbar.

The obtained value is one order of magnitude higher than those displayed by the perovskite compounds (NH$_4$)MF$_3$ (M$^{2+}$: Mn$^{2+}$, Co$^{2+}$, Cd$^{2+}$, Mg$^{2+}$ and Zn$^{2+}$) that also experience a 3-fold order-disorder phase transition due to ordering of the ammonium cations below $T_o$,[34] reflecting the more flexible nature of this formate compound as compared to fluorides.

This large pressure dependence means that upon application of adequate external pressures (> 24 kbar) and/or appropriated chemical substitutions that give rise to internal pressure it would be feasible to shift this transition to room temperature, making this and related materials much more interesting for technological applications.

**Conclusions**

We have carried out detailed calorimetric studies, thermal expansion measurements and powder as well as single crystal X-ray diffraction experiments on the hybrid inorganic-organic compound $[(CH_3)_2NH_2][Mn(HCOO)_3]$.

The obtained results directly proof the presence of an abrupt structural phase transition at $T_t \sim 187$ K, the temperature at which the sharp dielectric transition takes place. This structural phase transition, which involves a symmetry change from space group $R$-$3c$ to $Cc$, a discontinuous variation of the cell parameters and cell volume, is clearly of first order. These structural changes, which are observed by temperature dependent powder and single crystal X-ray diffraction, are also in good agreement with the data obtained by high-resolution thermal expansion measurements that show a very sharp and rather large step-like variation in the relative length change at a temperature very close to $T_t$. In addition, the calorimetric results clearly point out to a 3-fold order-disorder transition, as also supported by the crystal structure studies.

Concomitant to the transformation from $R$-$3c$ to $Cc$ twinning of the crystals occurs, and the LT-phase contains 3 twin domains related to each other by a rotation of ~120º around the $c^*$-axis, each of them corresponding in fact to one of the three possible orientations of the nitrogen atom of the DMA cations in the HT-phase. This result supports the mechanism of cooperative freezing of the molecular rotation of the DMA cations as origin of the associated dielectric transition.

From the determination of the crystal structure at temperatures close to the phase transition, we have obtained relevant information about the evolution of the [Mn(HCOO)$_3$]$^-$ framework and the DMA cations across the transition temperature. These results reveal the relevance of the distortion of the rather flexible [Mn(HCOO)$_3$]$^-$ framework, in particular the tilting of the [MnO$_6$] octahedra, in the overall structural transition that also involves the cooperative ordering of the DMA-cations (responsible in turn for the dielectric transition). It remains to be unravelled what mechanism is initially responsible for the transition, i.e. structural and/or dielectric transition.

Finally, based on the calorimetric and single crystal X-ray diffraction data and using the Clausius-Clapeyron equation, we have estimated the pressure dependence of the transition temperature in this compound. It turns to be rather large (d$T_t$/d$P$= 4.6 ± 0.1 K/kbar), as expected in view of the rather flexible structure of this MOF material, implying that upon application of adequate external or internal pressure it should be feasible to shift this transition to room temperature, rendering this and related materials much more interesting for technological applications.

Another very interesting aspect that remains to be explored is whether this parameter could also lead to enhancement of the magnetoelectric coupling in MOFs in the paramagnetic state near room temperature. This may well be the case due to increased magnetoelastic effects that favour both the superexchange interactions and the ferroelectric state.

**Experimental Section**

**Synthesis.** Single crystals of [(CH$_3$)$_2$NH$_2$][Mn(HCOO)$_3$] were obtained using the solvothermal route described in detail in reference [10], which at variance to other preparation methods described in the literature, [11,13] allows the synthesis of this compound in one single step. The as-obtained single crystals are small but suitable for studies such as single crystal X-ray diffraction. Nevertheless the fact that they are too small for physical measurements such as thermal expansion prompted as to obtain larger single crystals of [(CH$_3$)$_2$NH$_2$][Mn(HCOO)$_3$]. For this purpose, small crystals obtained by the initial procedure were used as seeds, and they were allow to grow from the mother solution at room temperature for about one month until single crystals of almost 1 mm$^3$ size were obtained. The crystals are colorless and typically show pseudo-cubic morphologies.

**Thermal analysis.** Differential scanning calorimetric (DSC) analyses were carried out in a TA Instruments MDSC Q-2000 modulated differential scanning calorimeter equipped with a liquid nitrogen cooling system. The temperature calibration was verified making use of the toluene melting point.

For these experiments, 11-12 mg of sample were placed in hermetic aluminium crucibles and subjected to several thermal cycles, that consisted of a linear cooling step from 220 to 125 K, followed by a heating ramp at the same rate, from 125 to 220 K. Heating/cooling rates of 5, 10, 20, and 40 K/min were used.

**Powder X-ray diffraction.** Room temperature powder X-ray diffraction (PXRD) data from crushed crystals were initially collected in a Siemens D-5000 diffractometer using CuKα radiation (λ= 1.5418 Å). X-ray scattering experiments as a function of temperature were performed in a Siemens D-500 diffractometer using CuKα radiation (λ= 1.5418 Å) equipped with a low-temperature helium closed-cycle refrigerator. The powder diffraction data were collected in the 2Θ-range 12º to 42º at 12 different temperatures between 205 K and 150 K.

The obtained PXRD patterns were analyzed by Le Bail profile analysis using the Rietica software.[35] The peak shapes were described by a pseudo-Voigt function and the background was modelled with a 6-term polynomial function.

**Single crystal X-ray diffraction.** Single-crystal data sets of the sample were collected at different temperatures, both on cooling and warming, between 220 and 150 K, in steps of 5, 2 and even 1 degree close to the transition temperature, in a Bruker Kappa Apex II X-ray diffractometer equipped with a CCD detector and using monochromatic MoKα radiation (λ=0.71073 Å). A suitable crystal of approximately 0.21 x 0.28 x 0.60 mm$^3$ was chosen and mounted on a MiTeGen MicroMount$^{TM}$ using FOMBLIN® YR-1800 perfluoro-polyether (Lancaster Synthesis). To carry out these experiments, the crystal was cooled at different rates using a cold stream of nitrogen from an Oxford Cryosystem cooler. The data integration and reduction was performed using the APEX2 V2013.2-0 (Bruker AXS, 2013) suite software. The intensity collected was corrected for Lorentz and polarization effects and for absorption by semi-empirical methods on the basis of symmetry-equivalent data using SADABS 2012/1 of the suite software. The structures were solved by the direct method using the SHELXS-97[36] program and were refined by least squares method on SHELXL-97.[37] The presence of multiple lattices was clear from visual inspection of diffraction images collected below the $T_t$ temperature. This dataset was indexed using CELL_NOW 2008/4 obtaining three solutions that interpret all the diffraction peaks with I/σ (I)>8 found in the 983 images measured. The integration of the reflections was performed with SAINT 8.30C taking into account the orientation matrices of the three twin domains simultaneously.

TWINABS 2012/1 was used for scaling and merging the twinned data. All the software used to treat those twinned data is included in the APEX2 suite.

To refine the structures, anisotropic thermal factors were employed for the non-H atoms. For $T > T_t$ the hydrogen atoms of the formate ions were found in the Fourier map and their coordinates and isotropic thermal factors were refined. However, the H-atoms of the DMA could not be found or introduced at idealized positions, due to the disordered arrangement of this cation.

Meanwhile for $T < T_t$, the hydrogen atoms of the $NH_2^+$ group of the DMA cations were found in the Fourier map and the rest of the hydrogen atoms of the DMA were introduced at idealized positions. All hydrogen atoms were restrained using the riding model implemented in SHELXL-97.

Pseudo precession photographs were reconstructed pixel by pixel from the complete sets of collected single-crystal X-ray diffraction images.

**Thermal expansion.** Thermal expansion experiments were carried out by employing a high-resolution capacitive dilatometer,[38] enabling to detect length changes $\Delta l \geq 10^{-2}$ Å. Data were collected in the temperature range 4 K $\leq T \leq$ 200 K. A single crystal of a suitable size (1.00 x 0.80 x 0.75 mm$^3$) was used for these studies, where the data were obtained on warming at rate of ~3 K/hour.

The crystal was oriented using a Laue Back-Reflection X-ray facility equipped with an image plate to detect X-rays.

**Electronic Supplementary Information (ESI) available:**

Table with calorimetric data, Table with selected bond angles, LeBail refinement of powder X-ray diffraction pattern at T=200 K and T=160 K, Evolution of the cell parameters and volume as a function of temperature, Single crystal images at different temperatures. Pseudo-precession images generated from single-crystal X-ray diffraction data. Scheme of the relative orientation of the twin domains. Cif files with the single crystal structure at 185 K and 190 K.

**Acknowledgements**

The authors are grateful for financial support from Ministerio de Economía y Competitividad (MINECO) (Spain) and EU under projects FEDER MAT2010-21342-C02-01, from Xunta de Galicia under project PGIDIT10PXB103272PR and from the German Science Foundation DFG in frame of SFB/TR49. L.C.G.A. also wants to thank UDC for a Predoctoral fellowship.

**FIGURE CAPTIONS**

Figure 1. Conventional perovskite structure view and unit cell of [(CH$_3$)$_2$NH$_2$][Mn(HCOO)$_3$] **(a):** at room temperature (HT trigonal phase) **(b):** at 100 K (LT monoclinic phase). In both pictures selected planes of interest are also indicated. As it can be seen the DMA cations is at the center of a ReO$_3$-type cavity, formed by manganese and formate ions. In the HT-phase, the N is disordered over three positions (Fig. 1a), while at LT-phase it sits in a single crystallographic position (Fig. 1b). In this latter the dash lines indicate H-bonds between the N atom and O atoms of formate.

Figure 2. DSC results as a function of temperature obtained by heating and cooling the [(CH$_3$)$_2$NH$_2$][Mn(HCOO)$_3$] sample at different temperature rates (5-40 K/min).

Figure 3. Powder X-ray diffraction (PXRD) patterns of [(CH$_3$)$_2$NH$_2$][Mn(HCOO)$_3$] obtained at different temperatures between 205 K and 150 K in steps of 5 K. Inset: Enlarged view of the x-ray diffraction pattern of [(CH$_3$)$_2$NH$_2$][Mn(HCOO)$_3$].

Figure 4. Temperature dependence of the cell parameters a, b and c and angle β expressed in terms of the monoclinic unit cell.

Figure 5. Variation of the relative length change as a function of temperature as obtained from thermal expansion measurements. Inset: Image of the single crystal of [(CH$_3$)$_2$NH$_2$][Mn(HCOO)$_3$] inside the thermal expansion cell.

Figure 6. Pseudo precession images generated from single-crystal X-ray diffraction data obtained at 190 K (HT-phase) and 185 K (LT-phase) displaying the reciprocal planes [0kl] and [hk0].

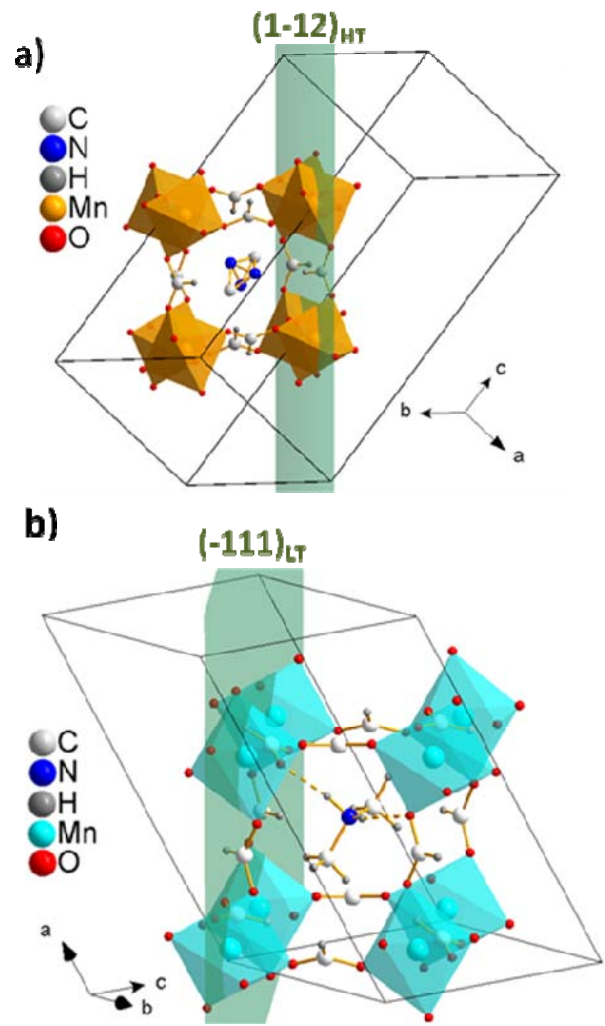

Fig. 1

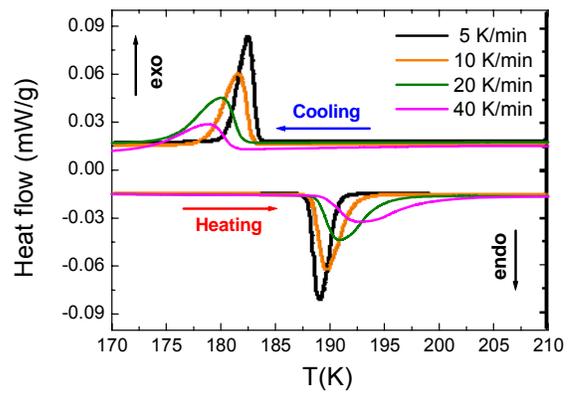

Fig. 2

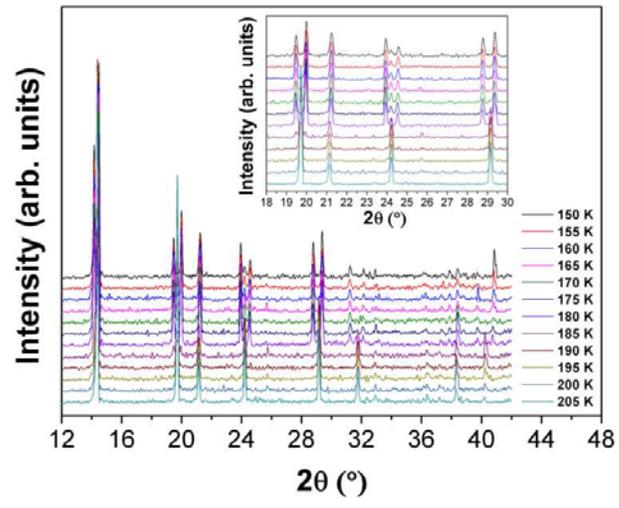

Fig. 3

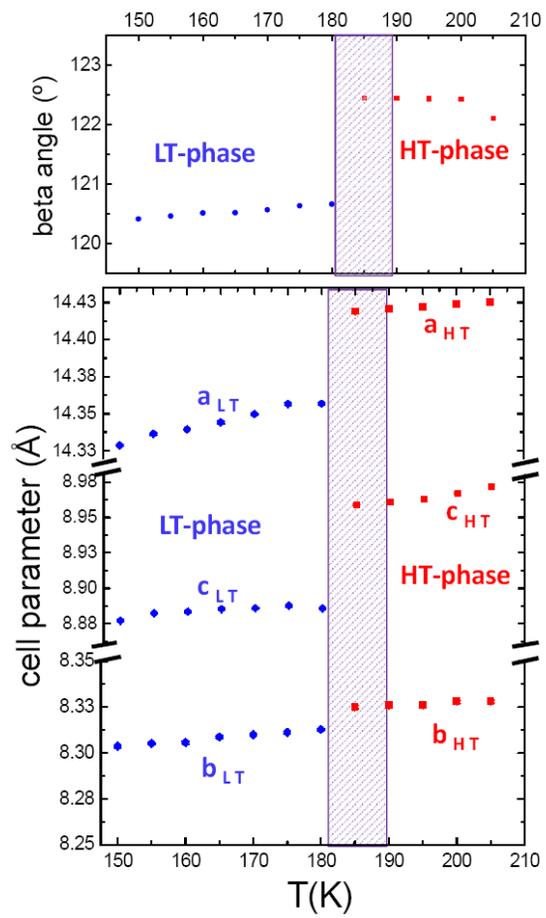

Fig. 4

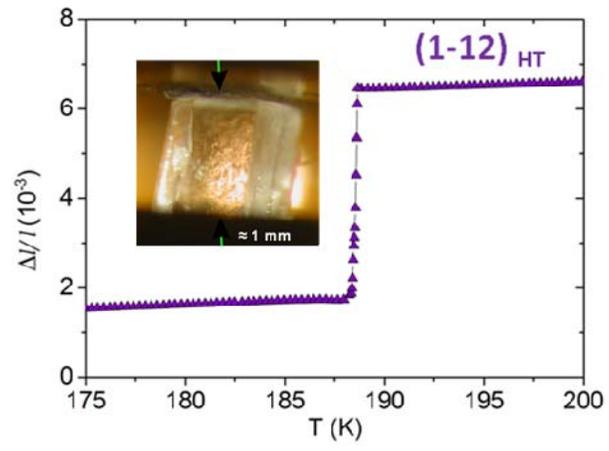

Fig. 5

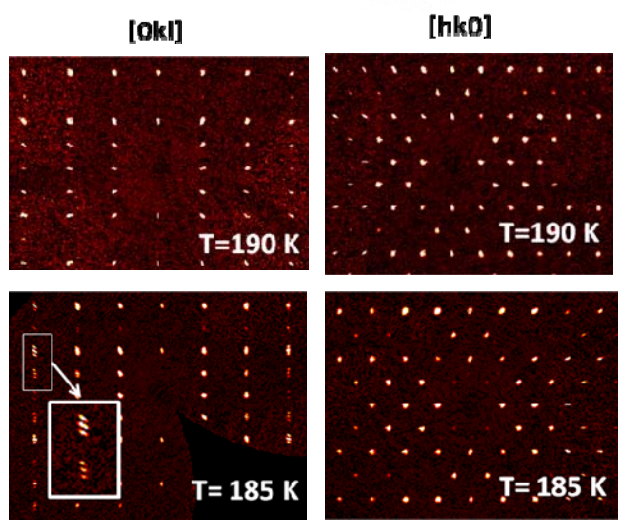

Fig. 6

# TABLE

Table I: Cell parameters, selected bond lengths, octahedral tilting and R-values corresponding to the HT (T=190 K) and LT phase (T=185 K) of [(CH$_3$)$_2$NH$_2$][Mn(HCOO)$_3$] just above and below the structural phase transition.

| T=190 K | | T=185K | |
|---|---|---|---|
| Cell parameters | | | |
| a (Å) | 8.3544(5) | a (Å) | 14.3644(17) |
| c (Å) | 22.8701(1) | b (Å) | 8.3180(10) |
| | | c (Å) | 8.8927(11) |
| | | β (°) | 120.750(5) |
| Mn environment | | | |
| Mn-O1 (Å) | 2.194 (1) | Mn-O1 (Å) | 2.182(6) |
| | | Mn-O2 (Å) | 2.173(6) |
| | | Mn-O3 (Å) | 2.167(7) |
| | | Mn-O4 (Å) | 2.168(6) |
| | | Mn-O5 (Å) | 2.204(6) |
| | | Mn-O6 (Å) | 2.202(6) |
| | | Mn-O$_{Average}$ (Å) | 2.183(6) |
| Formate | | | |
| C1-O1 (Å) | 1.249 (3) | C1-O1 (Å) | 1.250(13) |
| | | C1-O6 (Å) | 1.247(13) |
| | | C3-O2 (Å) | 1.228(12) |
| | | C3-O3 (Å) | 1.236(12) |
| | | C4-O4 (Å) | 1.242(11) |
| | | C4-O5 (Å) | 1.261(12) |
| | | C-O$_{Average}$ (Å) | 1.244(12) |
| DMA | | | |
| C2-N1 (Å) | 1.444 (8) | C21-N1 (Å) | 1.533(15) |
| | | C22-N1 (Å) | 1.463(14) |
| N-O $_{(H-bond)}$ (Å) | 2.908(11) | N-O $_{(H-bond)}$ (Å) | 2.835(11)/2.841(8) |
| Octahedra tilt (°) | | | |
| | 26.69(2) | | 27.2(2) |
| | 26.69(2) | | 27.4(2) |
| | 26.69(2) | | 24.1(2) |
| R-values | | | |
| Final $R_1$ values (I > 2σ(I)) | 0.0371 | Final $R_1$ values (I > 2σ(I)) | 0.0634 |
| Final $wR(F^2)$ values (I > 2σ(I)) | 0.1036 | Final $wR(F^2)$ values (I > 2σ(I)) | 0.1535 |
| Final $R_1$ values (all data) | 0.0426 | Final $R_1$ values (all data) | 0.1006 |
| Final $wR(F^2)$ values (all data) | 0.1066 | Final $wR(F^2)$ values (all data) | 0.1739 |
| Goodness of fit on $F^2$ | 1.200 | Goodness of fit on $F^2$ | 0.955 |

**Graphical and textual abstract**

Sharp and large step-like variation in the relative length change of a single crystal of the multiferroic MOF material [(CH$_3$)$_2$NH$_2$][M(HCOO)$_3$] that experiences a first-order structural phase transition at $T_t$~187 K, associated to its ferroelectric-paraelectric transition.

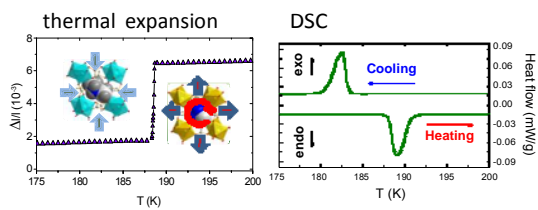

ELECTRONIC SUPPLEMENTARY INFORMATION

Table S1: Calorimetric characteristics of the transition in [(CH$_3$)$_2$NH$_2$][Mn(HCOO)$_3$].

| Heating/cooling rate (K/min) | T$_{max}$(K) heating/endo | ΔH (J/mol) heating/endo | ΔS (J/molK) heating/endo | T$_{max}$(K) cooling/exo | ΔT$_{heat-cool}$ | ΔH (J/mol) cooling/exo | ΔS (J/molK) cooling/exo |
|---|---|---|---|---|---|---|---|
| 5 | 189.17 | 1671.6 | 8.83 | 182.55 | 6.62 | 1701.6 | 9.32 |
| 10 | 189.84 | 1679.4 | 8.84 | 181.65 | 8.19 | 1713.4 | 9.43 |
| 20 | 191.08 | 1667.9 | 8.73 | 180.16 | 10.92 | 1710.6 | 9.49 |

It should be noted that DSC curves were also obtained on cooling and heating at 40 K/min. Nevertheless as those rates were not consistently kept along the corresponding ramps, the corresponding data are neither included in this Table nor were they considered for the calculation of ΔH$_{av}$, ΔS$_{av}$ and N in the DSC section.

Table S2.- Selected bond angles corresponding to the HT (T=190 K) and LT phase (T=185 K) of [(CH$_3$)$_2$NH$_2$][Mn(HCOO)$_3$] just above and below the structural phase transition

| T=190 K | | T=185K | |
|---|---|---|---|
| Bond angles (º) | | | |
| Oi-Mn-Oj | 89.21(8)/ 90.79(8) | Oi-Mn-Oj | 89.8(3)/ 90.5(3)/ 90.3(3)/ 88.8(3)/ 93.0(3)/ 91.1(3)/ 87.1(3)/ 89.2(2)/ 90.5(2)/ 90.1(2) |
| Mn-O1-C1 | 126.4(2) | Mn-O1-C1 | 129.0(6) |
| | | Mn-O2-C3 | 128.8(6) |
| | | Mn-O3-C3 | 126.0(6) |
| | | Mn-O4-C4 | 126.6(6) |
| | | Mn-O5-C4 | 124.3(6) |
| | | Mn-O6-C1 | 127.9(6) |
| O1-C1-O1 | 126.2(4) | O1-C1-O6 | 124.8(5) |
| | | O2-C3-O3 | 127.7(9) |
| | | O4-C4-O5 | 126.0(8) |
| C2-N1-C2 | 118.1(8) | C21-N1-C23 | 110.5(8) |

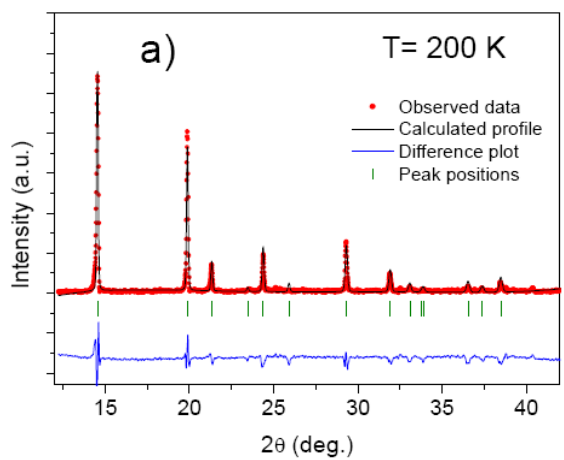

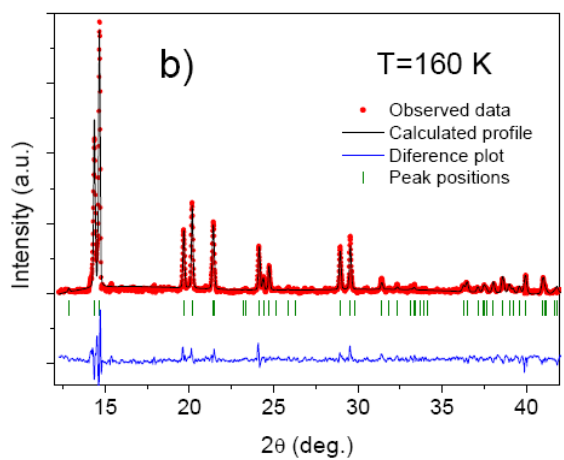

Figure S1.- LeBail refinement of powder X-ray diffraction pattern for [(CH$_3$)$_2$NH$_2$][Mn(HCOO)$_3$] compound at a) T=200 K (HT-phase) and b) T=160 K (LT-phase).

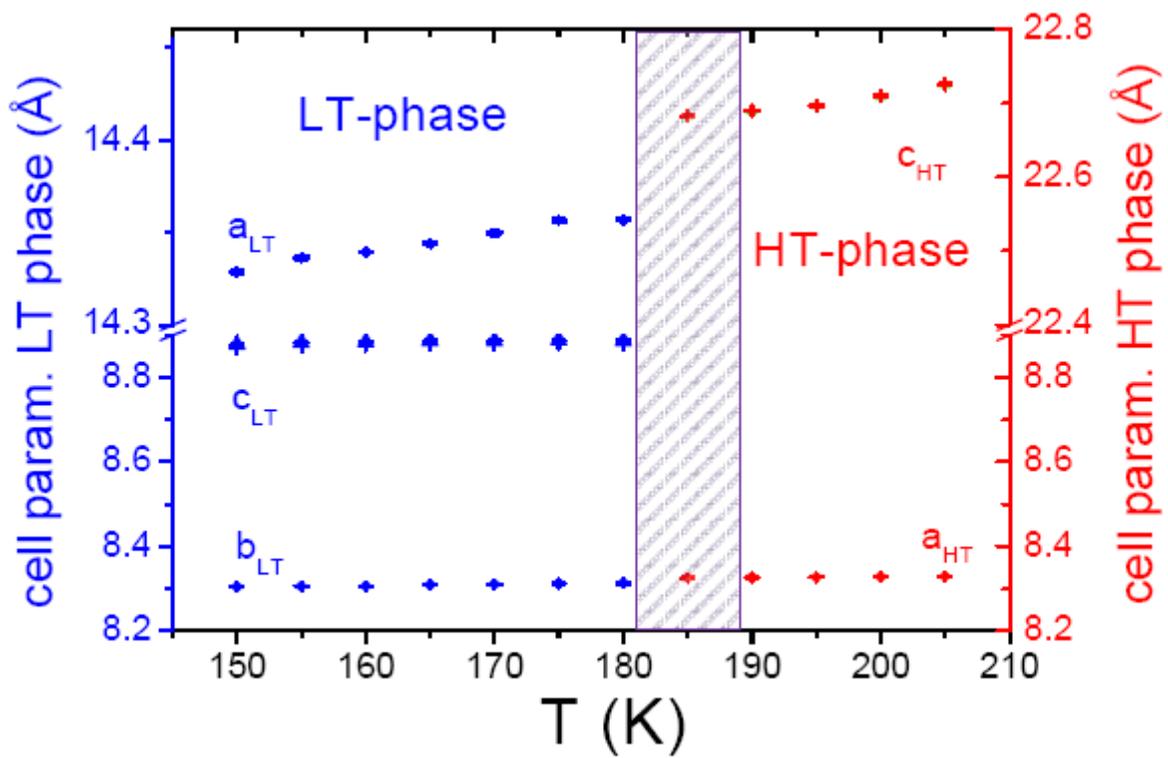

Fig. S2: Evolution of the cell parameters of the high temperature *R*-3*c* (Z=6) phase and of the low temperature *Cc* phase (Z=4) as a function of temperature.

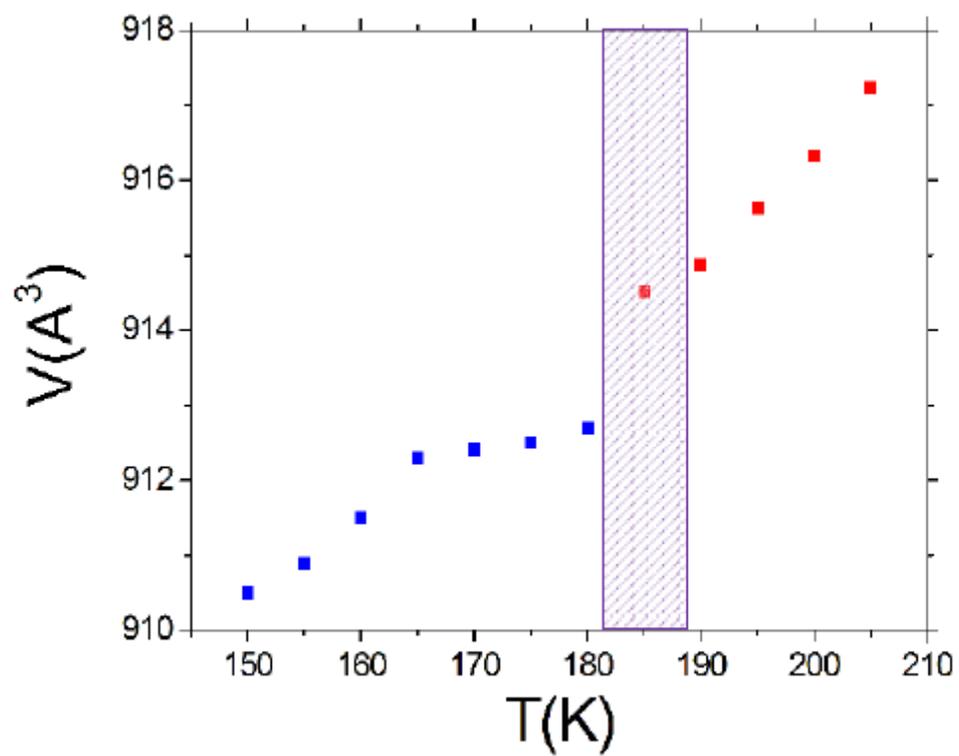

Fig. S3: Variation of the cell volume of $[(CH_3)_2NH_2][Mn(HCOO)_3]$ as a function of temperature.

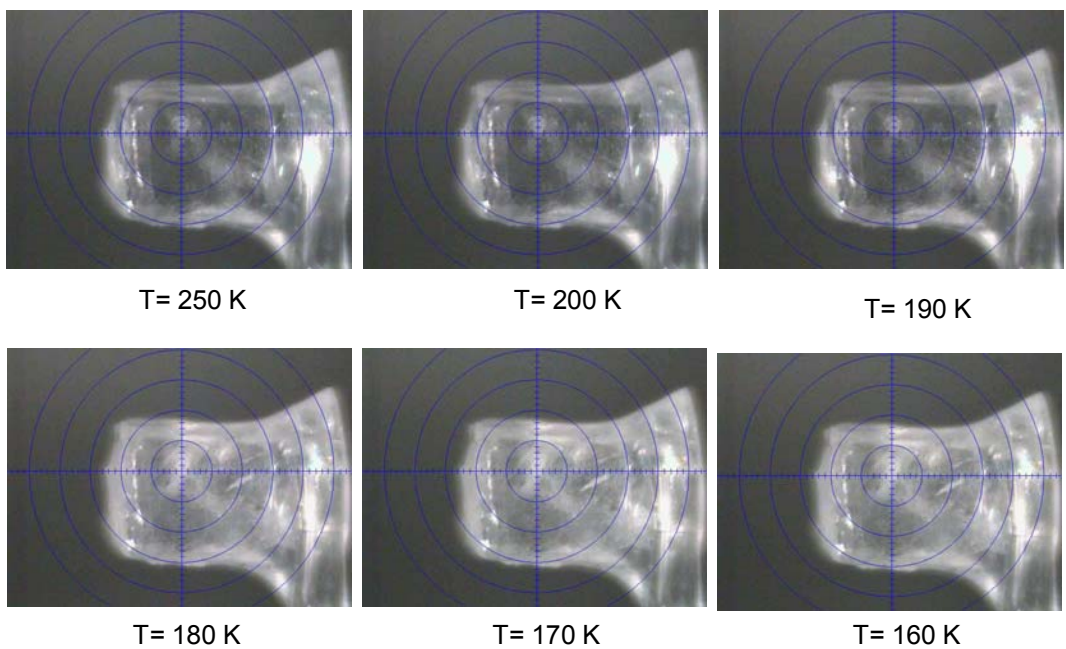

Fig. S4: Single crystal images of the [(CH$_3$)$_2$NH$_2$][Mn(HCOO)$_3$] compound at different temperatures. The pictures shows as the single crystal is completely transparent at T>185 K (HT-phase) and it becomes translucent at T<185 K (LT-phase). This process is completely reversible.

Figure S5: Pseudo precession images generated from single-crystal X-ray diffraction data obtained at 185 K displaying the a) [0kl] reciprocal plane, b) [h0l] and c) [hk0] respectively. In all cases the reciprocal axes shown correspond with the first (and most populated) twin domain of the monoclinic $Cc$ (LT) phase.

a)

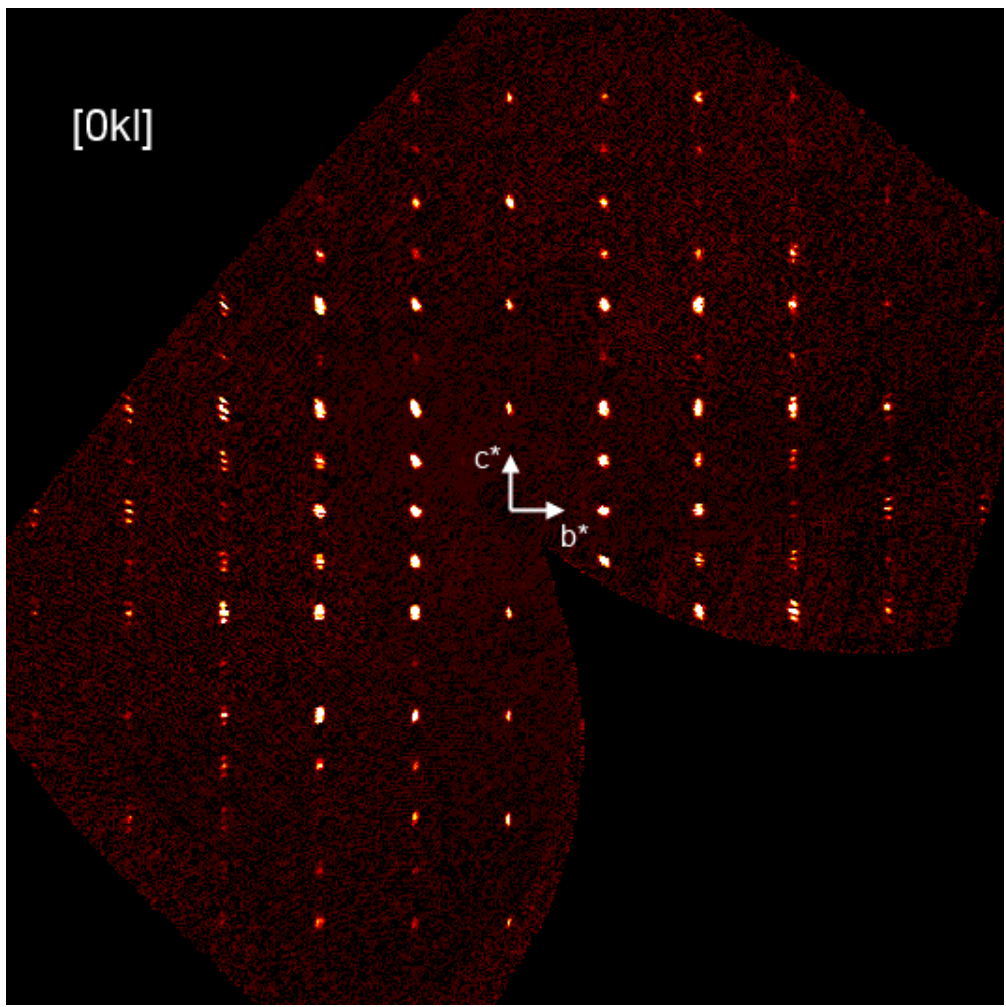

Zoom view with predicted reflections from twin domain 1, 2 and 3 respectively.

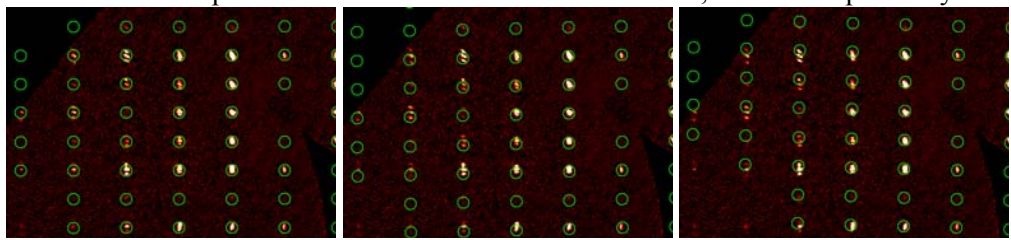

b)

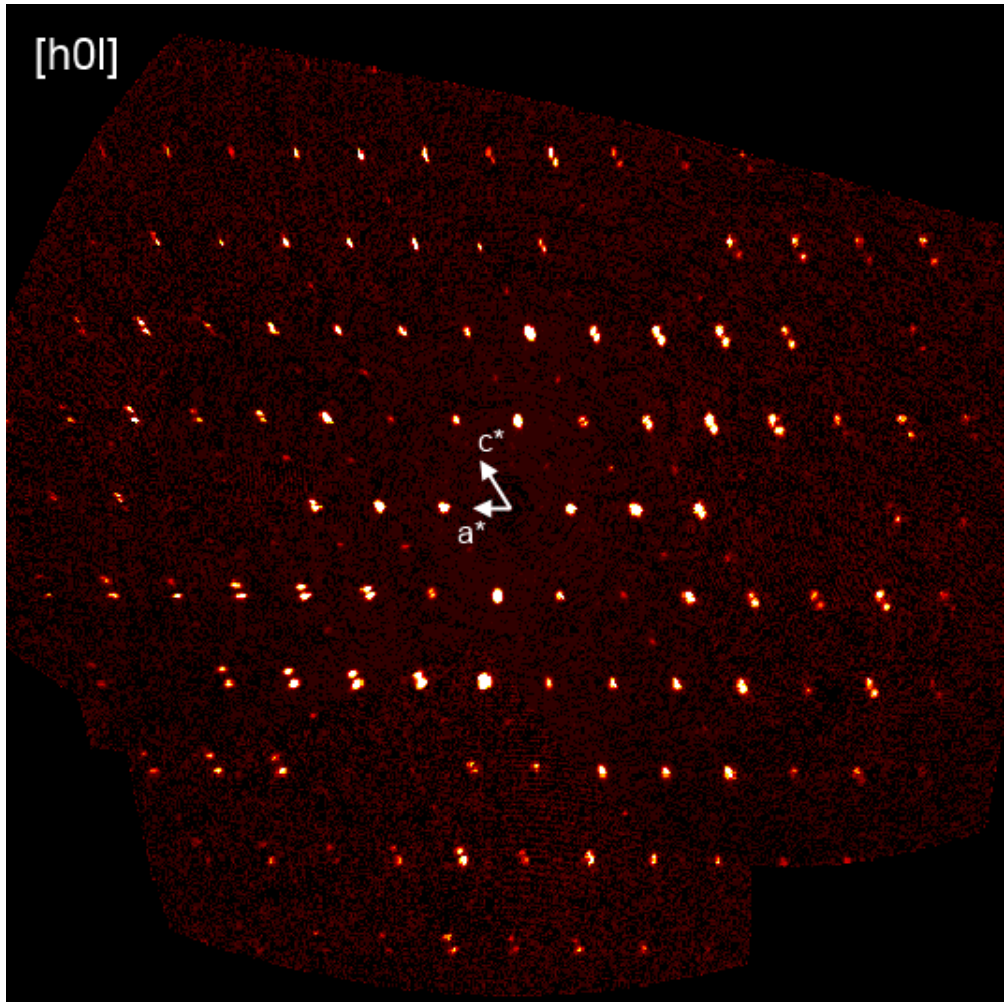

c)

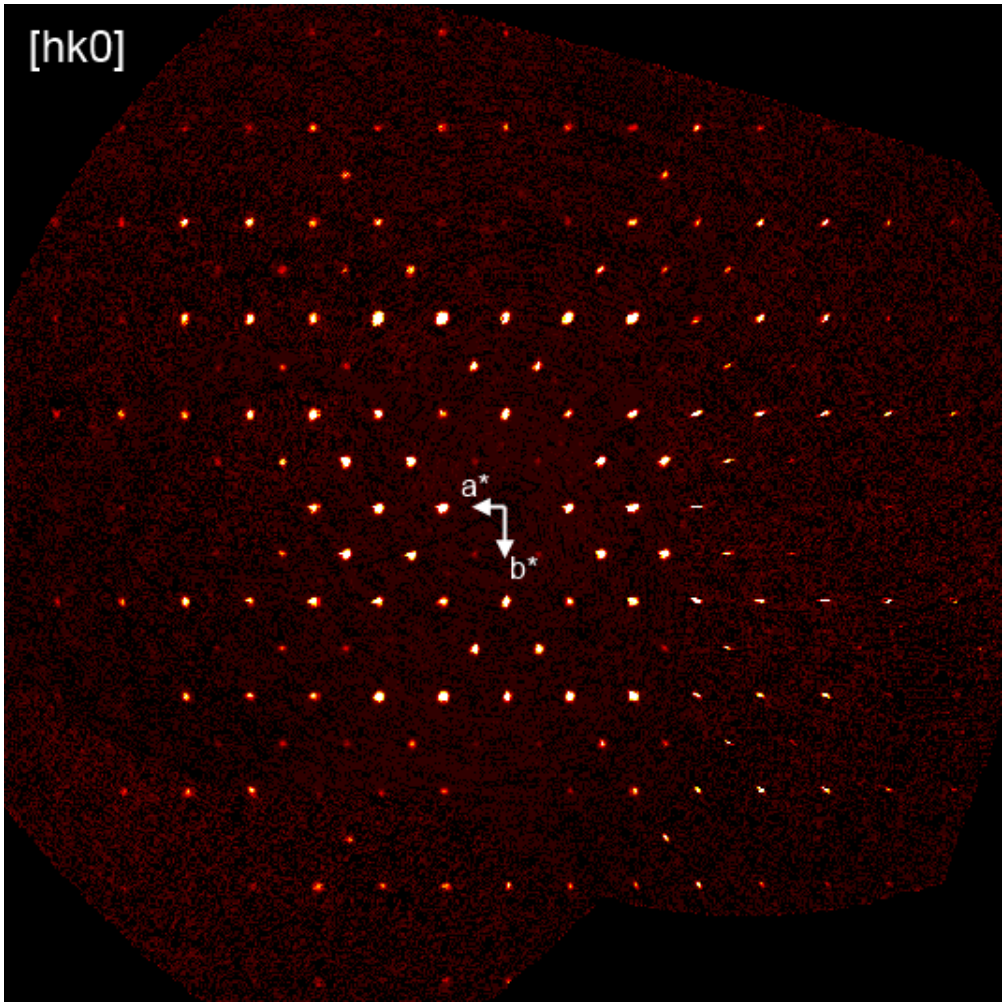

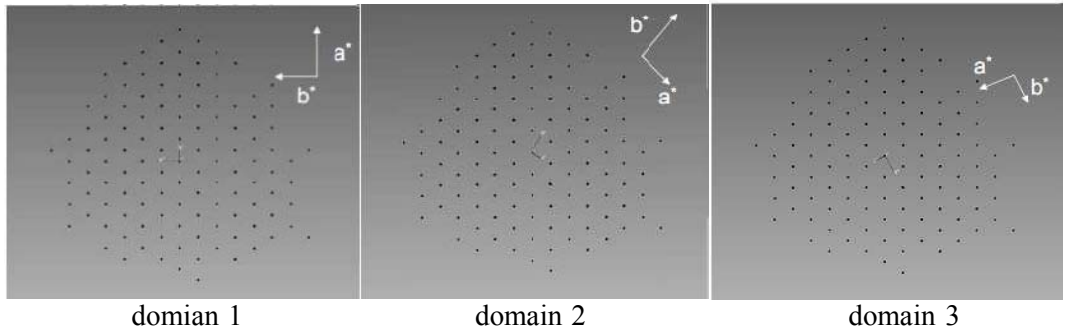

Fig. S6: Projections of the reciprocal space along the c* reciprocal axis direction of the data collected at 185K (monoclinic *Cc* structure). The relative orientations of the three twin domains found is also displayed. They are related by 120º rotations around the c* axis.

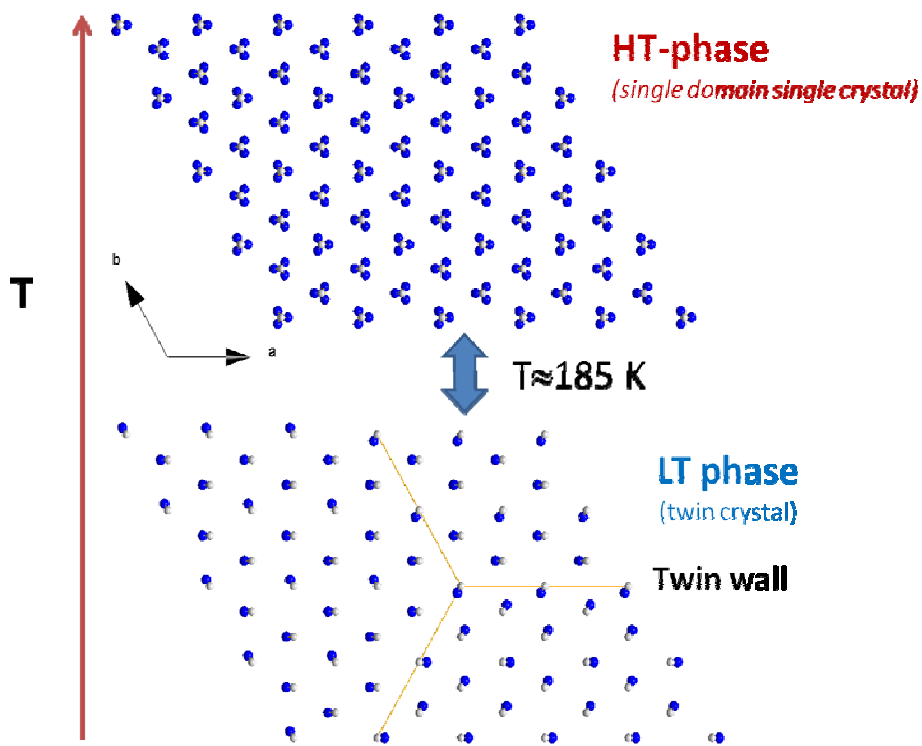

Fig. S7: Scheme of the relative orientation of the twin domains in the *Cc* monoclinic phase and its relationship to the high temperature *R*-3*c* trigonal phase. To facilitate its view, the [Mn(HCOO)$_3$]$^-$ framework of the structure has been removed and only the DMA cations of the structure are despicted. Blue circle (N-atoms) and white circle (C-atoms) of DMA cations.